\documentclass{icrc2009}

\usepackage{graphicx}   
\usepackage[font=footnotesize,caption=false]{subfig} 
\usepackage{fixltx2e}
\usepackage{url}

\newcommand{\shorttitle}[1]%
{\markboth{Proceedings of the 31\MakeLowercase{$^{st}$} ICRC, {\L}\'{o}d\'{z} 2009}{#1} }
\newcommand{\etal}{\MakeLowercase{\textit{et al. }}} 

\usepackage[latin1]{inputenc}

\begin{document}
\title{Search for High Energetic Neutrinos from Supernova Explosions with AMANDA}

\author{
\IEEEauthorblockN{Dirk Lennarz\IEEEauthorrefmark{1},
                  Jan-Patrick Hülß\IEEEauthorrefmark{1} and
                  Christopher Wiebusch\IEEEauthorrefmark{1}
                  for the IceCube Collaboration\IEEEauthorrefmark{2}
                  }
                  \\
\IEEEauthorblockA{\IEEEauthorrefmark{1}III. Physikalisches Institut, RWTH Aachen University, 52056 Aachen, Germany}
\IEEEauthorblockA{\IEEEauthorrefmark{2}See http://www.icecube.wisc.edu/collaboration/authorlists/2009/4.html for a full list of authors}
}

\shorttitle{Lennarz \etal AMANDA Supernova Search}
\maketitle

\begin{abstract}
Supernova explosions are among the most energetic phenomena in the known universe. There are suggestions that cosmic rays up to EeV energies might be accelerated in the young supernova shell on time scales of a few weeks to years, which would lead to TeV neutrino radiation. The data taken with the AMANDA neutrino telescope in the years 2000 to 2006 is analysed with a likelihood approach in order to search for directional and temporal coincidences between neutrino events and optically observed extra-galactic supernovae. The supernovae were stacked in order to enhance the sensitivity. A catalogue of relevant core-collapse supernovae has been created. This poster presents the results from the analysis.
\end{abstract}


\section{Introduction}
Almost a hundred years after their discovery, the acceleration mechanisms and sources of the cosmic rays remain an unsolved problem of modern astronomy. Neutrino astronomy can be an important contribution to the solution of this problem. Young supernovae in connection with a pulsar have been proposed as a possible source of cosmic rays with energies up to the ankle. This \emph{pulsar model} can be directly tested by measuring high energetic (TeV) neutrino radiation on time scales of a few weeks to years after the supernova \cite{pulsarmodel_1}\cite{pulsarmodel_3}.

The AMANDA-II neutrino telescope is located in the clear ice at the geographic South Pole and was fully operational since 2000. It reconstructs the direction of high energetic neutrinos by measuring Cherenkov light from secondary muons. The main background are muons and neutrinos produced in air showers in the atmosphere.

This analysis uses 7 years of AMANDA data taken during the years 2000-2006 with a total live-time of 1386 days. The data reconstruction and filtering is described in \cite{AMANDA_final_point_source_search} and the final event sample contains 6595 events. The contamination of mis-reconstructed atmospheric muon events is less than 5\% for a declination greater than $5^\circ$.

\section{Pulsar Model}
Rotational energy liberated by a pulsar can be converted into the energy of relativistic particles \cite{pulsarmodel_1}. Secondary particles, for example pions, are created in the interaction with the expanding supernova envelope and decay into neutrinos and other particles \cite{pulsarmodel_1}. In this analysis the pulsar model as described in \cite{pulsarmodel_3} is used. Thermonuclear supernovae have no pulsar inside the envelope and are therefore not considered by this model.

The phase of powerful, high energetic neutrino emission is limited by two characteristic times: the time at which the pion decay time becomes less than the time between two nuclear collisions ($t_\pi$) and the time at which the density of the envelope is sufficiently small for accelerated particles to escape into the interstellar space without interaction ($t_c$). The supernova neutrino luminosity as a function of time (\emph{model light curve}) is given by:
\setlength{\arraycolsep}{0.0em}
\begin{eqnarray}\label{eq:pulsar_model}
  L(t)&{}={}&\left(1-\exp\left(-\left(\frac{t_c}{t}\right)^2\right)\right)\cdot\frac{1}{1+\left(\frac{t_\pi}{t}\right)^3}\nonumber\\
  &&{\cdot}\:\lambda L_0\left(1+\frac{t}{\tau}\right)^{-2}\ ,
\end{eqnarray}\setlength{\arraycolsep}{5pt}where $\lambda$ is the fraction of the total magnetic dipole luminosity $L_0$ (in erg/s) that is transferred to accelerated particles and $\tau$ the characteristic pulsar braking time.

\begin{figure}[!t]
 \centering
 \includegraphics[width=0.515\textwidth]{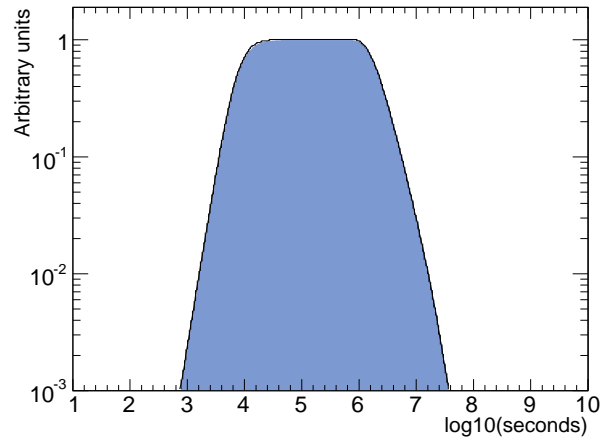}
 \caption{Typical supernova model neutrino light curve}
 \label{fig:Supernova_light_curve_typical}
\end{figure}
The shape and length of the model light curve depend on the supernova envelope mass ($\rm M_e$), uniformity (described by a parameter called $\xi$) and expansion velocity ($V$), the pulsar braking time and the maximum pion energy. An $E^{-2}$ neutrino energy spectrum is assumed with an energy cutoff at $10^{14}$ eV. Fig. \ref{fig:Supernova_light_curve_typical} shows a typical model light curve for $t_\pi\approx8\times10^3\rm s$ and $t_c\approx2\times10^6\rm s$. These values are obtained by choosing $\rm M_e = 3M_\odot$, $\xi = 1$, $V = 0.1\rm c$ and $\tau = 1$year.

\section{Supernova Catalogues}
For this analysis a catalogue of supernovae was created. It combines three different electronically available and regularly updated SN catalogues \cite{CBAT}\cite{ASC}\cite{SSC}. A comparison of the three catalogues revealed some inconsistencies in the listed information. A consistent selection was made with special attention to the objects mistaken for a supernova observation, the total number of supernovae and the supernova positions.

\begin{figure}[!ht]
  \centering
  \includegraphics[width=0.5\textwidth]{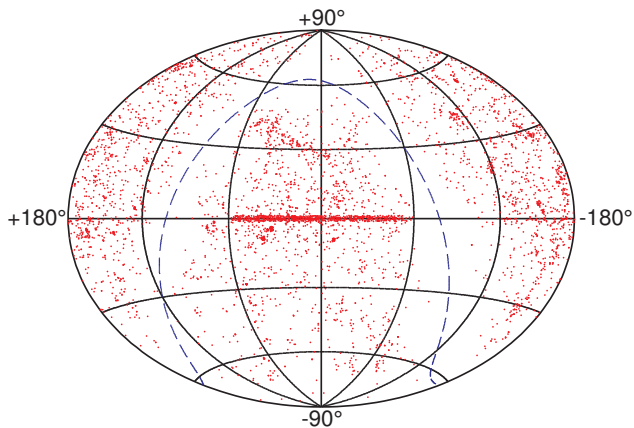}
  \caption{Distribution of observed supernovae in equatorial coordinates with the galactic plane indicated as dashed line. Due to the background from atmospheric muons only supernovae in the northern hemisphere are relevant.}
  \label{fig:Supernova_catalogues_SN_skymaps}
\end{figure}

Fig. \ref{fig:Supernova_catalogues_SN_skymaps} shows the distribution of the 4805 supernovae observed between 1885 and 2008. Out of these only about 700 supernovae inside the AMANDA data taking time are relevant. The clearly visible structure around the celestial equator are supernovae found by the Sloan Digital Sky Survey-II supernova survey. The nearest and best visible supernova for AMANDA was SN2004dj in NGC 2403 at a distance of approximately 3.33 Mpc

\begin{figure}[!ht]
  \centering
  \includegraphics[width=0.515\textwidth]{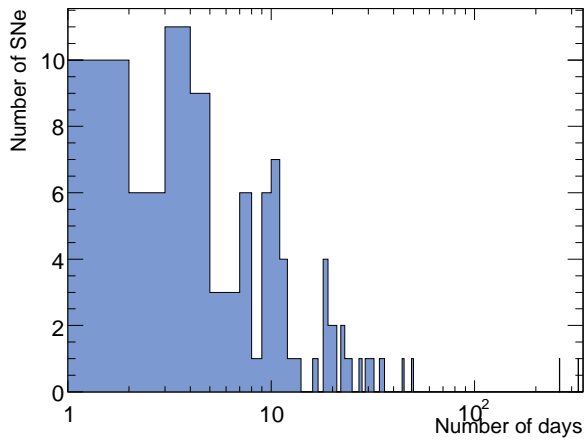}
  \caption{Number of days between the optical maximum and the discovery if the supernova was discovered after the maximum. A linear one day binning is shown on an logarithmic x-axis.}
  \label{fig:Supernova_catalogues_time_offset}
 \end{figure}

This analysis searches for directional and temporal coincidences between neutrinos and supernovae. Therefore additional input has to be quantified for each supernova. Firstly, the expected neutrino flux has to be determined from an accurate distance. The supernova distance can be identified with the distance to the host galaxy and can be estimated from the redshift. The redshift estimate is replaced by a measured distance (e.g. Cepheid variables or Tully-Fisher relation) if available. This improves the distance accuracy for nearby supernovae, which are most relevant.\\
Secondly, the explosion date is needed for the temporal correlation, but only the date of the optical maximum or the discovery date is available. From some well observed SNe (e.g. 1999ex and 2008D) it is known that the optical maximum occurs around 15-20 days after the explosion, which is used as a benchmark. Fig. \ref{fig:Supernova_catalogues_time_offset} shows the difference between the date of discovery and the date of maximum for those cases where the light curve was fitted to a template and the date of maximum extrapolated backwards in time or found on old photo plates. The majority of the supernovae are discovered within 20 days after the optical maximum. Hence, the discovery is assumed to be typically 20 days after the optical maximum. The uncertainty of the explosion date is accounted for in the likelihood approach (see next section).
 \\
Thirdly, for the individual supernova the needed input for the pulsar model is not available. Therefore, all supernovae are treated equally. The influence of the model light curve on the analysis is tested by defining two additional sets of parameters which result in light curves with very short and long neutrino emission. Hence, altogether three different model light curves (typical, short, long) are used. The width of the plateau that can be seen in Fig.  \ref{fig:Supernova_light_curve_typical} is 12 days for the typical, 1 day for the short and 76 days for the long light curve. The most realistic assumption for the supernovae in the catalogue is that they have individual realisations of the parameters of the pulsar model and therefore individual light curves between the extreme cases.

\section{Likelihood Approach}
A new likelihood approach was developed for this analysis \cite{Diplomarbeit}. Its principal idea is to compare all neutrino events from the experimental data sample to every relevant supernova and evaluate the likelihood ratio (LHR) between the hypothesis that this event is signal and the hypothesis of being background. This yields a large value for a good and small value for a bad match. The LHR for all events is summed in order to obtain a cumulative estimator, called $Q$:\setlength{\arraycolsep}{0.0em}
\begin{eqnarray}\label{eq:final_Q}
Q=\sum_\texttt{events} \frac{\sum_\texttt{SN}p(\vec{a}|\texttt{SN})p(\texttt{SN})}{p(\vec{a}|\texttt{BG})p(\texttt{BG})}\ ,
\end{eqnarray}\setlength{\arraycolsep}{5pt}where $\vec{a}$ are characteristic observables of the event.

The advantage of this likelihood definition is that it can be extended to a stacking analysis. $Q$ automatically assigns a small weight to irrelevant combinations of neutrinos and supernovae, while relevant ones receive a larger weight. Thus, all supernovae from the catalogue can be used in the analysis and no optimisation on the number of sources is needed. $Q$ is a sum of likelihood ratios and therefore its absolute value contains no physical information.

The probabilities in the likelihood sum are constructed from properties of AMANDA, the experimental data sample and the considered model light curve. $p(\texttt{BG})$ is the probability to have background and is an unknown but constant factor. This probability is eliminated by redefining $Q$ to $Q\cdot p(\texttt{BG})$.

$p(\vec{a}|\texttt{BG})$ is the probability that, assuming an event is background, it is observed at its specific time and from its specific direction. It is factorised into a temporal and an angular part. The temporal part corresponds to the AMANDA live-time. However, it cancels out with the corresponding temporal part of $p(\vec{a}|\texttt{SN})$. AMANDA does not distinguish between signal and background neutrinos and was obviously taking data when the event was measured. The angular probability is constructed with the normalised zenith angle distribution of the experimental data sample (see Fig. 2 in \cite{AMANDA_final_point_source_search}). The azimuth probability is constant, because AMANDA is completely rotated each day and the azimuth is randomised for the relevant time scales of this analysis.

The supernova signal probability consists of $p(\vec{a}|\texttt{SN})$ and $p(\texttt{SN})$. The first part depends on the specific event and is the probability that an event from a supernova is observed at a given time from a given direction. $p(\texttt{SN})$ is the probability to observe a signal from that supernova and is estimated for each supernova.

For $p(\vec{a}|\texttt{SN})$ two terms are considered. $p(\Psi|\texttt{SN})$ is the probability that a neutrino from a supernova is reconstructed with an angular difference $\Psi$ relative to the supernova direction. This probability is calculated from the point-spread function, which is obtained from Monte Carlo simulations. The second term $p(t,t_\texttt{SN}|\texttt{SN})$ yields the probability that a neutrino arrives with a time offset $t-t_\texttt{SN}$ from the explosion date. This probability is taken from a likelihood light curve.

\begin{figure}[!ht]
  \centering
  \includegraphics[width=0.515\textwidth]{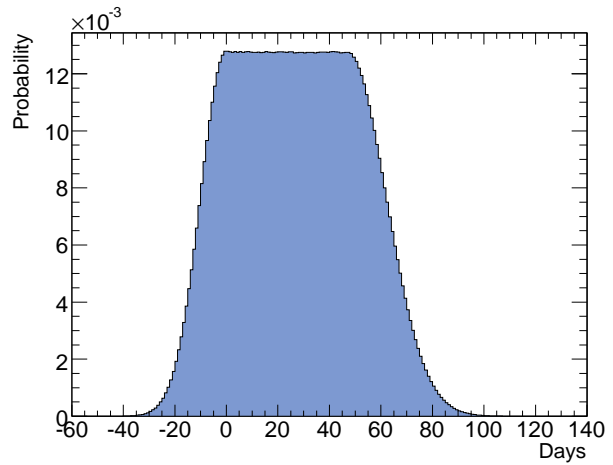}
  \caption{Typical likelihood light curve}
  \label{fig:Supernova_likelihood_light_curve_typical}
 \end{figure}

In order to be less model dependent three generic likelihood light curves (typical, short, long) are constructed. They are inspired by the model light curves and constructed conservatively in order to not miss signal by accidentally looking too early or too late. Hence, if the date of the optical maximum is known, the starting time for the likelihood light curves ($t=0$) is defined to be 30 days earlier. In case only the date of the discovery is known a 50 days earlier starting time is used. This makes sure that the explosion is not missed, because the time shift to the explosion date is overestimated by about 15 days for the optical maximum and up to 35 days for the date of discovery. The likelihood light curves consist of a half Gaussian for $t<0$, a plateau for $t>0$ and another half Gaussian after the plateau. The length of the plateau is the full width at 90\% of the model light curves and enlarged by the uncertainty of the explosion day. This uncertainty is bigger if only the date of discovery is known. The width of the Gaussian after the plateau is the full width at half maximum (FWHM) after the plateau of the model light curves. Fig. \ref{fig:Supernova_likelihood_light_curve_typical} shows the typical likelihood light curve for the date of discovery.

$p(\texttt{SN})$ depends on the supernova neutrino luminosity, distance and direction. The absolute value of $p(\texttt{SN})$ is determined by the supernova neutrino luminosity and is a free parameter of this analysis. However, the absolute normalisation is not required, because a constant factor results in a rescaling of $Q$ and hence only relative values are important. All supernovae are assumed to have the same neutrino luminosity at source. $p(\texttt{SN})$ decreases like the flux with the square supernova distance. AMANDA is not equally sensitive to neutrinos from all directions. Therefore the angular acceptance for different supernova directions is taken into account.

\section{Signal and Background Simulation}
$Q$ distributions for signal and background simulations are used to construct confidence belts with the Feldman-Cousins approach to the analysis of small signals \cite{Feldman_Cousins}. Each simulated data sample contains 6595 signal or background events like the experimental data set. Background events are simulated with the zenith angle distribution of the experimental data and the AMANDA live-time. For the signal simulation a model light curve and the AMANDA angular and temporal acceptance is simulated. The angular acceptance includes a random simulation of assumed systematic uncertainties of the measured rate of high energetic muon neutrinos \cite{AMANDA_final_point_source_search}. The simulation of the temporal and angular acceptance reduces signal events from days with low live-time or unfavourable supernova directions.

The confidence belts are used to estimate the sensitivity of the analysis. The sensitivity for the long model light curve is not compatible compared to \cite{AMANDA_final_point_source_search}. Furthermore, if the supernovae have short model light curves, the sensitivity is comparable for the short and typical likelihood light curves. The typical pulsar model is best detected with the typical likelihood light curve. Therefore the experimental data is analysed with the typical likelihood light curve, because it can cover a larger range of possible parameters.

\section{Experimental Result}
\begin{figure}[!ht]
  \centering
  \includegraphics[width=0.515\textwidth]{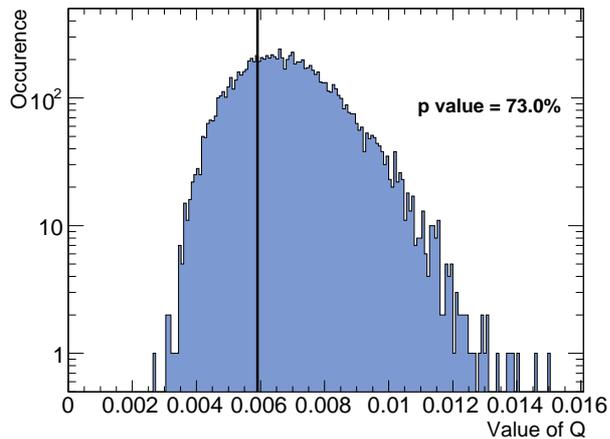}
  \caption{$Q$ value from analysing the experimental data sample with the typical likelihood light curve (horizontal line) in a distribution for background only.}
  \label{fig:Supernova_unblinding_Q_typical}
 \end{figure}

Analysing the experimental AMANDA data with the typical likelihood light curve yields:
 \setlength{\arraycolsep}{0.0em}
\begin{eqnarray}\label{Q_result}
 Q_{\rm typical}^{\rm Exp} = 0.0059 \ .
\end{eqnarray}
\setlength{\arraycolsep}{5pt}
Fig. \ref{fig:Supernova_unblinding_Q_typical} shows this value in a $Q$ distribution for background only. The p value of obtaining a $Q$ value equal or bigger than 0.0059 is 73.0\%. Hence, the $Q$ value is consistent with background and no deviation from the background only hypothesis is found. Therefore upper limits for the three model light curves are derived.

90\% upper limits on the signal strength are derived from the Feldman-Cousins confidence belts. With the help of the signal simulation explained above this value converts to the sum of neutrinos from all supernovae. Assuming the above model ranking of sources and the stacking this limit can also be interpreted as a limit on the number of neutrinos from SN2004dj. Tab. \ref{tab:UL} shows the obtained upper limits for the typical, short and long pulsar model light curve.
\begin{table}[!ht]
  \centering
  \begin{tabular}{|c|c|c|c|}
  \hline
  Pulsar model & All SNe  & SN2004dj \\
  \hline
  Typical      & $< 5.4$  & $< 1.0$\\
  \hline
  Short        & $< 4.1$  & $< 0.9$\\
  \hline
  Long         & $< 67.3$ & $< 5.9$\\
  \hline
  \end{tabular}
  \caption{90\% upper limits on the number of neutrinos from all supernovae and from SN2004dj.}
  \label{tab:UL}
\end{table}
\newpage
The event numbers can be converted to a flux by integrating the AMANDA neutrino effective area with the expected signal energy spectrum ($E^{-2}$ spectrum with cutoff at $10^{14}$ eV). Assuming the typical pulsar model for all supernovae and taking the average of the effective area over all directions, the 90\% upper limit on the flux from all supernovae for the plateau of powerful neutrino radiation (12 days) is:
\setlength{\arraycolsep}{0.0em}
\begin{eqnarray}\label{flux_limit}
  \frac{d\phi}{dE}\cdot E^2< 5.2\times10^{-6}\rm{\frac{GeV}{cm^2 s}}\ .
\end{eqnarray}\setlength{\arraycolsep}{5pt}

Using the effective area for the direction of SN2004dj, the corresponding 90\% upper limit for SN2004dj is:
\setlength{\arraycolsep}{0.0em}
\begin{eqnarray}\label{flux_limit}
  \frac{d\phi}{dE}\cdot E^2< 8.4\times10^{-7}\rm{\frac{GeV}{cm^2 s}}\ .
\end{eqnarray}\setlength{\arraycolsep}{5pt}

These limits are valid in the energy range from 1.1 TeV to 84.0 TeV.

Assuming that the energy range of the pulsar model as described in \cite{pulsarmodel_3} can be extended to higher energies, the limits would improve by about 30\% and are then valid in the energy range from 1.7 TeV to 2 PeV.

\section{Conclusion}
For the first time the neutrino emission from young supernova shells was experimentally investigated. In the context of the pulsar model no deviation from the background only hypothesis was found.

For a galactic supernova the expected flux from the pulsar model should be sufficient to be detectable by IceCube, the AMANDA successor. The sensitivity of this analysis might be enhanced by using an energy estimator in the likelihood and the individual event reconstruction error instead of the energy averaged point-spread function.

\end{document}